\documentclass[a4paper,12pt]{article}
\pdfoutput=1
\usepackage{mathptmx} 
\usepackage{amsmath} 
\usepackage{amsfonts} 
\usepackage{graphicx} 
\usepackage{hyperref} 
\usepackage[format=hang,justification=justified,labelfont=bf,labelsep=quad]{caption} 
\def\mm#1{\makebox(0,0){\strut#1}} 
\def\lm#1{\makebox(0,0)[l]{\strut#1}}

\def\Gammaprime{\Gamma\kern.07em'}
\newtheorem{theorem}{Theorem}
\newtheorem{proposition}{Proposition}
\newtheorem{corollary}{Corollary}
\newcommand{\ttp}[1]{^{\kern.07em#1}} 
\def\proof{\noindent{\em Proof.\enspace}}
\def\endproof{\hfill\strut\nobreak\hfill\tombstone\par\medbreak}
\def\tombstone{\hbox{\lower.4pt\vbox{\hrule\hbox{\vrule
  \kern7.6pt\vrule height7.6pt}\hrule}\kern.5pt}}
\newdimen\einr

\title{%
Recursive Inspection Games%
}
\author{Bernhard von Stengel%
\thanks{Department of Mathematics,
London School of Economics, London WC2A 2AE, United Kingdom.
Email: stengel@nash.lse.ac.uk}
}
\date{\normalsize February 7, 2016}

\begin{document}
\maketitle

\begin{abstract}
We consider a sequential inspection game where an inspector
uses a limited number of inspections over a larger number of
time periods to detect a violation (an illegal act) of an
inspectee.
Compared with earlier models, we allow varying rewards
to the inspectee for successful violations.
As one possible example, the most valuable reward may be the
completion of a sequence of thefts of nuclear material
needed to build a nuclear bomb.
The inspectee can observe the inspector, but the inspector
can only determine if a violation happens during a stage
where he inspects, which terminates the game; otherwise the
game continues.

Under reasonable assumptions for the payoffs, the
inspector's strategy is independent of the number of
successful violations.
This allows to apply a recursive description of the game,
even though this normally assumes fully informed players
after each stage.
The resulting recursive equation in three variables for the
equilibrium payoff of the game, which generalizes several
other known equations of this kind, is solved explicitly
in terms of sums of binomial coefficients.

We also extend this approach to non-zero-sum games and,
similar to Maschler (1966), ``inspector leadership'' where
the inspector commits to (the same) randomized inspection
schedule, but the inspectee acts legally (rather than mixes
as in the simultaneous game) as long as inspections remain.  

\vskip4ex


\vskip1ex
\noindent
\textit{MSC2010 subject classification:}
91A05, 
91A20  

\vskip1ex
\noindent
\textit{OR/MS subject classification:}
Games/group decisions: Noncooperative;
Military: Search/surveillance


\vskip2ex

\noindent
Published in: \textit{Mathematics of Operations Research}
(2016),\\
\url{http://dx.doi.org/10.1287/moor.2015.0762}

\end{abstract}

\maketitle

\newpage
\section{Introduction}
\label{s-intro}

Inspection games model situations where an \emph{inspector}
with limited resources verifies by means of
\emph{inspections} that an \emph{inspectee} adheres to legal
obligations, which the inspectee has an incentive to
\emph{violate}.
Inspection games have been applied to arms control and
disarmament, tax auditing, fare evasion, environmental
pollution, and homeland security; for a survey see Avenhaus,
von Stengel, and Zamir \cite{AvvSZa2002}, 
Avenhaus and Canty \cite{AvCa2012},
and other recent works such as
\cite{De2013} or \cite{DF2014}.


This paper presents a generalization of a classical
sequential inspection game by Dresher \cite{Dr1962}.
In Dresher's game, the inspector has to distribute a given
number of inspections over a larger number of inspection
periods to detect a violation that the inspectee,
who can count the inspector's visits, performs in at most
one of these periods.
In our extension of this game, the inspectee may violate
more than once, and collect a possibly different reward for
each successful violation; for example, a violation may be
the diversion of a certain amount of nuclear material in
a time period, with the highest reward to the inspectee
once he has diverted enough material to build a nuclear
bomb.
As in Dresher's game, the game ends if a violation is
discovered by the inspector who inspects at the same time.
This is in line with an application to arms control, and may
also apply in other contexts where an identified violator of
legal rules becomes subject to much tighter surveillance.

A central aspect of our model and its analysis, and the
reason for its choice of parameters, is that the inspector's
mixed strategy in equilibrium does not depend on whether a
successful violation took place during a time period without
an inspection, about which the inspector is normally not
informed.
As we will explain in \S\ref{s-strategy}, the game can
therefore, despite this lack of information, be described
recursively by a sequence of $2\times2$ games for each
stage.
As long as there are remaining inspections (but fewer than
the number of remaining time periods) and intended
violations, the inspector and inspectee randomize at each
stage whether to inspect and to violate.  
For the payoffs and mixed strategy probabilities in
equilibrium we give explicit solutions in terms of the game
parameters.

Our analysis starts with a zero-sum game, which is then
extended to non-zero-sum payoffs.
Furthermore, if, as in Maschler \cite{Ma1966}, the inspector can
\emph{commit} to his mixed equilibrium strategy, the
inspectee will act legally as long as inspections remain.
This commitment power, known as ``inspector leadership'',
increases the inspector's payoff.

The main precursor to this work is 
\cite{vS1991},
which, however, only considers the extra parameter of a
varying number of intended violations, not different rewards
for them.
Inspection games with two parameters (time periods and
inspections) were considered by Dresher \cite{Dr1962}, Thomas and
Nisgav \cite{ThNi1976}, and Baston and Bostock \cite{BaBo1991}.
Maschler \cite{Ma1966} introduced non-zero-sum games and inspector
leadership.
H\"opfinger \cite{Ho1971} and Avenhaus and von Stengel
\cite{AvvS1992}
extended Dresher's model to non-zero-sum payoffs. 
Rinderle \cite{Ri1996} studied the case that inspections may
have probabilities of false alarms and non-detection of a
violation.
Avenhaus and Canty \cite{AvCa2005} considered sequential inspections
where timeliness of detection matters.
Games with a third parameter of intended violations were
considered by Kuhn \cite{Ku1963}, Sakaguchi \cite{Sa1977, Sa1994}, and
Ferguson and Melolidakis \cite{FeMe1998}.
In these models, the game continues even after a detected
violation, unlike in our model.
In addition, the inspector is fully informed after each
stage whether a violation took place or not even when he did
not inspect.
As already noted by Kuhn \cite[p.~174]{Ku1963}, this full
information is implicit in a recursive description.  

In \S\ref{s-model}, we describe the inspection game
and its parameters.
The recursive equation for the value of this zero-sum game
is solved explicitly.
The equilibrium strategy of the inspector depends only on
the number of remaining time periods and inspections. 
Section~\ref{s-strategy} discusses the key property that the
strategy of the inspector does not depend on the inspectee's
intended violations and their rewards, which
allows to apply the solution also when the inspector has no
information about undiscovered violations.
We also show that our model is as general as possible to
achieve this property.
In \S\ref{s-nonzero}, we show how to extend the
solution relatively easily to the non-zero-sum game where a
detected violation incurs a negative cost to both players
compared to the case of legal action and no inspection.
The ``inspector leadership'' game is studied in
\S\ref{s-lead}.
In \S\ref{s-conclude}, we discuss
possible extensions of our model, 
and general aspects of the recursive games we consider,
in particular computational advantages compared to games in
extensive form.

\section{Zero-sum inspection game with multiple violations}
\label{s-model}

We consider a two-player game $\Gamma(n,m,k)$,
were $n,m,k$ are three nonnegative integer parameters.
The game is played over $n$ discrete \emph{time periods}.  
The number $m$ is the number of \emph{inspections} available
to the inspector (the first player).
The number $k$ is the maximum number of \emph{intended
violations} of the inspectee (the second player).
In each time period, the inspector can use one of his
inspections (if $m>0$) or not, and simultaneously (if $k>0$)
the inspectee chooses between legal action and violation.
The game has also a real-valued ``penalty'' parameter $b$
and nonnegative ``reward'' parameters
$r_k,r_{k-1},\ldots,r_1$ that determine the payoffs, as
follows.

In this section, we assume that the payoffs are zero-sum.
Let $v(n,m,k)$ be the value of the game $\Gamma(n,m,k)$, as
the equilibrium payoff to the inspector.
If $n=0$, then the game is over and $v(n,m,k)=0$.
More generally, if $m\ge n$, then the inspector can inspect
in every remaining time period, where we assume that the
inspectee acts legally throughout, with 
\begin{equation}
\label{mn}
v(n,m,k)=0\qquad
\hbox{if }m\ge n.
\end{equation}
If $n>0$, the game $\Gamma(n,m,k)$ is described recursively.
Suppose first that $n>m>0$, so that the inspector decides
whether to use one of his inspections or not, and $k>0$, so
the inspectee decides whether to act legally or to violate.
The recursive description of $\Gamma(n,m,k)$, with value
$v(n,m,k)$, is given by the following payoffs to the
inspector, which are the costs to the inspectee, at the
first time period.
\begin{equation}
\label{rec}
\setlength{\unitlength}{1.2em}
\lower3.3\unitlength\hbox{%
\begin{picture}(25,7)(0,-.5)
\put(0,6){\line(0,-1){6}}
\put(6,6){\line(0,-1){6}}
\put(25,6){\line(0,-1){6}}
\put(15,6){\line(0,-1){6}}
\put(0,2){\line(1,0){25}}
\put(0,6){\line(3,-1){6}}
\put(0,6){\line(1,0){25}}
\put(0,4){\line(1,0){25}}
\put(0,0){\line(1,0){25}}
\put(1.9,4.5){\mm{{inspector}}}
\put(4.1,5.5){\mm{{inspectee}}}
\put(1,3){\lm{{inspection}}}
\put(7.5,3){\lm{$v(n-1,{m-1},k)$}}
\put(20,3){\mm{$b\cdot{r_k}$}}
\put(1,1){\lm{{no inspection}}}
\put(7.5,1){\lm{$v(n-1,m,k)$}}
\put(20,1){\mm{$v(n-1,m,{k-1})-{r_k}$}}
\put(10.25,5){\mm{{legal action}}}
\put(20,5){\mm{{violation}}}
\end{picture}} 
\end{equation}
Of the four possible combinations of the player's actions in
the first period, one of them \emph{terminates} the game,
namely when the inspector inspects and the inspectee
performs a violation, which we assume is caught with
certainty.
In all other cases, the game continues.
If the inspectee acts legally and the inspector inspects,
then the game continues as the game $\Gamma(n-1,m-1,k)$.
If the inspectee acts legally and the inspector does not
inspect, then the game continues as the game
$\Gamma(n-1,m,k)$.
If the inspectee violates and the inspector did not inspect,
then the game continues as the game ${\Gamma(n-1,m,k-1)}$,
where in addition the inspectee collects the \emph{reward}
$r_k$ which he can keep even if he is caught in a later time
period. 
The corresponding bottom-right cell in (\ref{rec})
has therefore payoff entry $v(n-1,m,k-1)-r_k$ to the
inspector.

If the game terminates because the inspectee is caught, we
assume that inspectee has to pay the \emph{penalty}
$b\cdot r_k$\,, which is proportional to his reward $r_k$ if
the violation had been successful,
multiplied by the penalty factor~$b$.
We assume only that $b>-1$, to allow for the
possibility (in particular when payoffs are no longer
zero-sum, discussed in \S\ref{s-nonzero}) that even a
caught violation is less preferred by the inspector than
legal action (with reference payoff zero).
A single successful (uncaught) violation with payoff $-r_k$
should however still be worse for the inspector than a
caught violation with payoff $b\cdot r_k$, hence the
requirement that $b>-1$.
This condition holds obviously when $b>0$ where a caught
violation creates an actual penalty to the inspectee that is
worse than legal action.

The nonnegative rewards to the inspectee
$r_k,r_{k-1},\ldots,r_1$ are numbered in that order 
to identify them from the game parameter $k$ in
$\Gamma(n,m,k)$ as the game progresses.
That is, $r_k$ is the reward for the first successful
violation, $r_{k-1}$ for the second, and so on until the
reward $r_1$ for the $k$th and last violation if the
game has not ended earlier.
The inspectee can perform at most one violation per time
period.
Hence, if there are no inspections left ($m=0$), then the
inspectee can violate in each of the remaining $n$ time
periods up to $k$ times in total, that is, 
\begin{equation}
\label{m0}
v(n,0,k)=
-\sum_{i=1}^{\min\{k,n\}}
r_{k+1-i}~.
\end{equation}
We allow some rewards to be zero.
If all remaining rewards are zero, then this gives the same
payoffs as when the inspectee only acts legally from now on,
so this may instead be represented by a smaller~$k$.
However, the term $v(n,0,k)$ in (\ref{m0}) may be zero even
if some remaining rewards are nonzero.
This case may arise in the course of the game after some
time periods without violations so that $n<k$, for example
when $n=1$, $k=2$, $r_2=0$, $r_1=1$, so we allow for this
possibility.

The game $\Gamma(n,m,k)$ is completely described by the
``base cases'' (\ref{mn}) and (\ref{m0}) (both of which
imply $v(0,m,k)=0$) and the recursive description
(\ref{rec}).
This description of the game assumes that both players are
fully informed about the other player's action after each
time period, and thus know in which of the four cells in
(\ref{rec}) the game continues.
We call this the game with \emph{full information} and will
weaken this assumption in \S\ref{s-strategy}.

The following main theorem gives an explicit formula for the
game value $v(n,m,k)$ and the optimal inspection strategy.
A large part of the proof is to show that (\ref{rec}) has a
circular preference structure and hence a mixed equilibrium.
The most important, but very direct part of the proof (from
(\ref{eq}) onwards) is that the explicit representation
(\ref{v}) holds.
We discuss a possible derivation of the term $t(n,m,k)$ in
\S\ref{s-lead} after Theorem~\ref{t-com}.

\begin{theorem}
\label{t-main} 
Let $n,m,k$ be nonnegative integers,
$b>-1$, and $r_k,r_{k-1},\cdots,r_1\ge 0$.
Define
\begin{equation}
\label{snm}
s(n,m)=\sum_{i=0}^m{n\choose i}b^{m-i}
\end{equation}
and
\begin{equation}
\label{t}
t(n,m,k)=\sum_{i=1}^kr_{k+1-i}{n-i\choose m}.
\end{equation}
Then the zero-sum game $\Gamma(n,m,k)$ defined by
$(\ref{rec})$ for $n>m>0$ and $k>0$ and by $(\ref{mn})$ and
$(\ref{m0})$ otherwise has value 
\begin{equation}
\label{v}
v(n,m,k)=\frac{-t(n,m,k)}{s(n,m)}~.
\end{equation}
For $n>m>0$ and $k>0$\,, the game $(\ref{rec})$ has a
completely mixed equilibrium where the inspector inspects
with probability $p$ and the inspectee violates with
probability $q$, where 
\begin{equation}
\label{p}
p=\frac{s(n-1,m-1)}{s(n,m)}~,
\qquad 1-p=\frac{s(n-1,m)}{s(n,m)}~,
\end{equation}
and
\begin{equation}
\label{q}
q=\frac{v(n-1,m,k)-v(n-1,m-1,k)}
{v(n-1,m,k)-v(n-1,m-1,k)+b\cdot r_k-v(n-1,m,k-1)+r_k}~.
\end{equation}
This equilibrium is unique, unless $r_{k+1-i}=0$ for
$1\le i\le\min\{k,n-m\}$, in which case all entries in
$(\ref{rec})$ are zero and the players can play arbitrarily.  
\end{theorem}

\proof{Proof.}
We first consider some properties of $s(n,m)$ as defined in
(\ref{snm}). 
Clearly,
\begin{equation}
\label{s0}
s(n,0)=1, \qquad
s(n,n)=(1+b)^n
\end{equation}
and
\begin{equation}
\label{sb}
b\cdot s(n-1,m-1)= \sum_{i=0}^{m-1}{n-1\choose i}b^{m-i}
=s(n-1,m)-{n-1\choose m}.
\end{equation}
Furthermore,
\begin{equation}
\label{ssum}
s(n,m)=s(n-1,m-1)+s(n-1,m)
\qquad(0<m<n)
\end{equation}
which holds because
\begin{equation}
\label{proofssum}
\begin{array}{rcl}
s(n,m)
&=&\displaystyle
\sum_{i=0}^m{n\choose i}b^{m-i}
~=~
{n\choose 0}b^m +\sum_{i=1}^m\bigl({n-1\choose
i}+{n-1\choose i-1}\bigr)b^{m-i}
\\&=&\displaystyle 
{n-1\choose 0}b^m+\sum_{i=1}^m{n-1\choose i}b^{m-i} 
   +\sum_{i=0}^{m-1}{n-1\choose i}b^{m-1-i}\\
\\&=&s(n-1,m)+s(n-1,m-1)\,.
\end{array}
\end{equation}
This means $s(n,m)$ is uniquely defined inductively by
(\ref{s0}) and~(\ref{ssum}).
Recall that ${x\choose 0}=1$ for any $x$
\cite[p.~50]{Fe1968}.
The following alternative representation 
\begin{equation}
\label{salt}
s(n,m)=\sum_{i=0}^m{n-1-i\choose m-i}(1+b)^{i} 
\end{equation}
holds because it also fulfills (\ref{s0})
and (\ref{ssum}), which is
shown similarly to (\ref{proofssum}).
Because $b>-1$, we have $s(n,m)>0$ for $0\le m\le n$ by
(\ref{salt}) (or by (\ref{s0}) and~(\ref{ssum})).

The main assertion to prove is the explicit representation
(\ref{v}) for $v(n,m,k)$.
Clearly, $v(n,m,0)=0$, and (\ref{m0}) and (\ref{mn})
hold because in (\ref{t}), ${n-i\choose m}=0$ if $i>n-m$.
Hence, we can assume that $n>m>0$ and $k>0$ where the
recursive description (\ref{rec}) applies.
By induction on $n$, we can assume as inductive hypothesis that
\begin{equation}
\label{ACD}
 A={v(n-1,m-1,k)},
 \qquad
 C=v(n-1,m,k), 
 \qquad
 D=v(n-1,m,k-1)-r_k
\end{equation}
are given using~(\ref{v}). 
These numbers and $B=b\cdot r_k$ define the game (\ref{rec})
as 
\begin{equation}
\label{ABCD}
\renewcommand\arraystretch{1.5}
\begin{array}{r|c|c|}
\multicolumn{1}{c}{} & \multicolumn{1}{c}{1-q} & \multicolumn{1}{c}{q}
\\
\cline{2-3}
p~~& ~~~~A~~~~ & ~~~~B~~~~\\
\cline{2-3}
1-p~~& C & D\\ 
\cline{2-3}
\end{array}
\end{equation}
which also shows the probabilities $p$, $1-p$ and $1-q$, $q$
of playing the rows and columns.
To complete the induction, we will show that this game has
value $v(n,m,k)$ as in (\ref{v}).

If $r_{k+1-i}=0$ for $1\le i\le\min\{k,n-m\}$, then by
(\ref{t}) and (\ref{v}) $A=B=C=D=0$, so this is the all-zero
game with value zero in agreement with (\ref{v}), and
arbitrary equilibrium strategies of the players.
So assume that this is not the case, so that 
\begin{equation}
\label{tpos}
t(n-1,m-1,k)>0
\end{equation}
and hence $A<0$.

Intuitively, (\ref{rec}) has a mixed equilibrium because the
inspector prefers not to inspect if the inspectee acts
legally and to inspect if he violates, and the inspectee
prefers to act legally if inspected and to violate
otherwise.
In (\ref{ABCD}), this holds if
\begin{equation}
\label{ineq}
A<C, \quad B>D,\quad
A<B, \quad C>D~.
\quad
\end{equation}
It is easy to see (and well known) that then with
\begin{equation}
\label{pp}
p=\frac{C-D}{B-A+C-D}~,
\qquad
1-p=\frac{B-A}{B-A+C-D}~,
\end{equation}
\begin{equation}
\label{qq}
1-q=\frac{B-D}{C-A+B-D}~,
\qquad
q=\frac{C-A}{C-A+B-D}~,
\end{equation} 
the game has value $v(n,m,k)$, where
\begin{eqnarray}
v(n,m,k)&=& p\cdot A+ (1-p)\cdot C
~=~ p\cdot B+ (1-p)\cdot D~,
\label{eep}
\\
\label{eeq}
v(n,m,k)&=& (1-q)\cdot A+ q\cdot B\kern.2em
~=~ (1-q)\cdot C+ q\cdot D~,
\end{eqnarray}
where $p$ in (\ref{pp}) is uniquely determined by (\ref{eep})
and $q$ in (\ref{qq}) is uniquely determined by (\ref{eeq}),
and $p$, $1-p$, $q$, and $1-q$ are all positive by (\ref{ineq}).

For (\ref{ineq}), we first show $A<C$.  
By (\ref{v}), this is equivalent to 
\[
\frac{-t(n-1,m-1,k)}{s(n-1,m-1)}
<
\frac{-t(n-1,m,k)}{s(n-1,m)}
\]
or, by (\ref{tpos}), to
\begin{equation}
\label{ssr}
\frac{s(n-1,m)}{s(n-1,m-1)}
>
\frac{t(n-1,m,k)}{t(n-1,m-1,k)} 
=
\frac{
r_k{n-2\choose m}+r_{k-1}{n-3\choose
m}+\cdots+r_{1}{n-1-k\choose m}
}{
r_k{n-2\choose m-1}+r_{k-1}{n-3\choose
m-1}+\cdots+r_{1}{n-1-k\choose m-1} 
}~. 
\end{equation}
Assume that $k\le n-m$, otherwise replace $k$ by $n-m$
because ${n-1-i\choose m-1}={n-1-i\choose m}=0$ for $i>n-m$.
We show that the right expression in (\ref{ssr}) is largest
when $r_k>0$ and $r_{k-1}=\cdots=r_1=0$.
Namely, for general nonnegative $\rho_1,\ldots,\rho_k$, not
all zero, positive $h_1,\ldots,h_k$, and any
$g_1,\ldots,g_k$ so that 
\begin{equation}
\label{ghgiven}
\frac{g_1}{h_1}\ge
\frac{g_2}{h_2}\ge
\cdots\ge
\frac{g_k}{h_k}~,
\end{equation} 
we have
\begin{equation}
\label{gh}
\frac{g_1}{h_1}\ge 
\frac{\rho_1g_1+\cdots+\rho_kg_k}
{\rho_1h_1+\cdots+\rho_kh_k}
\end{equation}
which is seen by induction as follows.
By omitting the terms where $\rho_i=0$,
we can assume $\rho_i>0$ for all~$i$.
For $k=1$, (\ref{gh}) is true.
For $k>1$, let $G=\rho_1g_1+\rho_2g_2$ and
$H=\rho_1h_1+\rho_2h_2$.
Then
\begin{equation}
\label{GH}
\frac{g_1}{h_1}\ge
\frac{G}{H}=
\frac{\rho_1g_1+\rho_2g_2}{\rho_1h_1+\rho_2h_2}
\ge
\frac{g_2}{h_2}
\end{equation}
because the left inequality in (\ref{GH}) is equivalent to
$g_1(\rho_1h_1+\rho_2h_2)\ge h_1(\rho_1g_1+\rho_2g_2)$
and thus to 
$g_1\rho_2h_2\ge h_1\rho_2g_2$
which holds by (\ref{ghgiven}); 
the right inequality in (\ref{GH}) is shown similarly.
This shows (\ref{gh}) for $k=2$, and for $k>2$ using the
inductive hypothesis 
\[
\frac{G}{H}\ge
\frac{g_3}{h_3}\ge
\cdots\ge
\frac{g_k}{h_k}
\qquad\Rightarrow\qquad
\frac{G}{H}\ge
\frac{G+\rho_3g_3+\cdots+\rho_kg_k}
{H+\rho_3h_3+\cdots+\rho_kh_k}~.
\]
With $\rho_i=r_{k+1-i}$\,,
$g_i={n-1-i\choose m}$,
$h_i={n-1-i\choose m-1}$ for $1\le i\le k$
we have $\frac{g_i}{h_i}=\frac{n-i-m}m$ and thus
(\ref{ghgiven}) and (\ref{gh}), so (\ref{ssr}) holds if
\begin{equation}
\label{AC}
\frac{s(n-1,m)}{s(n-1,m-1)}
>
\frac{{n-2\choose m}}{{n-2\choose m-1}}
=\frac{n-1-m}m
\end{equation}
which we now show.
By (\ref{salt}), the following are equivalent:
\[
\renewcommand\arraystretch{1.5}
\begin{array}{rcl}
s(n-1,m)
&>&
\displaystyle
\frac{n-1-m}{m}\cdot s(n-1,m-1)~,
\\
\displaystyle
\sum_{i=0}^{m}{n-1-i\choose m-i}(1+b)^{i}
&>&
\displaystyle
\frac{n-1-m}{m}\cdot
\sum_{i=0}^{m-1}{n-1-i\choose m-1-i}(1+b)^{i}~,
\\
\displaystyle
(1+b)^{m}~+~\sum_{i=0}^{m-1}{n-1-i\choose m-1-i}\frac{n-m}{m-i}(1+b)^{i}
&>&
\displaystyle
\sum_{i=0}^{m-1}{n-1-i\choose m-1-i}
\frac{n-1-m}{m}
(1+b)^{i}~,
\end{array}
\]
which is true because $0<m<n$ and thus
$\frac{n-m}{m-i}> \frac{n-1-m}{m}$ for $0\le i<m$.
This shows (\ref{AC}) and thus $A<C$.

The remaining inequalities in (\ref{ineq}) are seen as
follows.
Because $b>-1$ and $r_k\ge0$, we have
$B=b\cdot r_k \ge 0-r_k\ge v(n-1,m,k-1)-r_k=D$,
with inequality possible only if $r_k=0$ and 
$t(n-1,m,k-1)=0$, which because
\begin{equation}
\label{tk}
t(n-1,m,k-1)~=~ 
r_{k-1}{n-2\choose m}+r_{k-2}{n-3\choose
m}+\cdots+r_{1}{n-k\choose m}
\end{equation}
means $r_{k+1-i}=0$ for $1\le i\le\min\{k,n-m\}$ which we
have excluded.
So $B>D$.
Suppose $p$ given by (\ref{p}) fulfills (\ref{eep}), which
implies~(\ref{pp}).
By (\ref{ssum}), the real number $p$ defined in (\ref{p}) is
indeed a probability.
Also, $p>0$ and $1-p>0$, so that by (\ref{pp}) either
$C<D$ and $B<A$ or $C>D$ and $B>A$.
The former can be excluded because it would imply
$B<A<C<D<B$.
This proves~(\ref{ineq}).

So it remains to show (\ref{eep}), that is, 
\begin{alignat}2
v(n,m,k) ={}&
p\cdot v(n-1,m-1,k)&{}+{}&(1-p)\cdot v(n-1,m,k)\,,
\label{eq}
\\
v(n,m,k) ={}&
p\cdot b\cdot r_k&{}+{}&(1-p)\cdot(v(n-1,m,k-1)-r_k)\,.
\label{eq2}
\end{alignat}
After multiplication with $s(n,m)$, (\ref{eq}) and
(\ref{eq2}) are by (\ref{v}) and (\ref{p}) equivalent to
\begin{alignat}2
{}-t(n,m,k)={}&
-\,t(n-1,m-1,k)&
~-~&t(n-1,m,k)\,,
\label{eqt1}
\\
{}-t(n,m,k)={}&
s(n-1,m-1)\cdot b\cdot r_k&
~+~&(-\,t(n-1,m,k-1)-s(n-1,m)\cdot r_k)\,.
\label{eqt2}
\end{alignat}
Equation (\ref{eqt1}) holds because, by (\ref{t}),
\renewcommand\arraystretch{1.5}
\begin{alignat}2
t(n-1,m-1,k)
+
t(n-1,m,k)
&=
\displaystyle
\sum_{i=1}^kr_{k+1-i}{n-1-i\choose m-1}
&{}+{}&
\displaystyle
\sum_{i=1}^kr_{k+1-i}{n-1-i\choose m}
\nonumber
\\
&=
\displaystyle
\sum_{i=1}^kr_{k+1-i}{n-i\choose m}
&{}={}&
t(n,m,k)\,.
\label{tt}
\end{alignat}
Equation (\ref{eqt2}) holds because, by (\ref{sb}) and (\ref{tk}),
\begin{equation}
\label{ttt}
\renewcommand\arraystretch{1.5}
\arraycolsep.2em
\begin{array}{rclcl}
&&s(n-1,m-1)\cdot b\cdot r_k&+&(-\,t(n-1,m,k-1)-s(n-1,m)\cdot r_k)
\\
&=&
(s(n-1,m)-{n-1\choose m})\cdot r_k&+&(-\,t(n-1,m,k-1)-s(n-1,m)\cdot r_k)
\\
&=&
{}-{n-1\choose m}\cdot r_k&-&t(n-1,m,k-1)
\\
&=&
{}-t(n,m,k)\,.  
\end{array}
\end{equation}
This shows (\ref{eq}) and (\ref{eq2}),
which completes the induction on~$n$.  

The inspectee's violation probability $q$ in (\ref{q}) is
just given by (\ref{qq}).
\endproof

By Theorem~\ref{t-main}, the game in (\ref{rec}) has a
unique mixed equilibrium (unless all payoffs are zero).
This uniqueness applies recursively to all stages of
$\Gamma(n,m,k)$ if the players use behavior strategies.
The same probabilities for their actions could result from
mixed strategies that correlate these actions, which we do
not consider because behavior strategies suffice
\cite{Ku1953}.

A special case of Theorem~\ref{t-main} has been shown 
in \cite{vS1991},
namely when $r_i=1$ for $k\ge i\ge1$.
In that case, $t(n,m,k)$ in (\ref{t}) can be written as
\begin{equation}
\label{r=1}
t(n,m,k)=\sum_{i=1}^k{n-i\choose m}=
{n\choose m+1}-{n-k\choose m+1}
\end{equation}
(see Feller \cite[p.~63, equation (12.6)]{Fe1968}).

Dresher \cite{Dr1962} actually considered two special cases of this
game for $b=1$, namely $k=1$ and $k=n-m$, where (\ref{r=1})
simplifies to 
\begin{equation}
\label{k1nm}
t(n,m,1)= {n-1\choose m}
\qquad
\hbox{and}
\qquad
t(n,m,n-m)= {n\choose m+1}.
\end{equation}
The corresponding expressions (\ref{v}) were stated and
proved by Dresher \cite{Dr1962}, and, apparently independently,
by Sakaguchi \cite{Sa1994}.

\section{Discussion of the model and interpretation of the
main theorem}
\label{s-strategy}

In this section, we discuss the main Theorem~\ref{t-main},
in particular the fact that the inspector's equilibrium
strategy depends only on the number of time periods and
inspections.
Consequently, the same strategy also applies to a new game
$\Gammaprime(n,m,k)$ where the inspector is not informed
about violations at previous time periods when he did not
inspect, which we call the game \emph{without full
information}.
In a basic form, this assumption is implicit in the models
by Dresher \cite{Dr1962}.

The recursive definition of $\Gamma(n,m,k)$ as in
(\ref{rec}) allows to compute the game value even without an
explicit formula as stated in (\ref{v}).
If a game as in (\ref{ABCD}) fulfills the inequalities
(\ref{ineq}) so that the game has a mixed equilibrium, then
the equilibrium probabilities (\ref{pp}) and~(\ref{qq}) give
the value of the game as
$\frac{BC-AD}{B-A+C-D}$.
Sakaguchi \cite{Sa1994} recursively computes the game value
$v(n,m,k)$ in this way for different entries in (\ref{rec}).

As mentioned, the recursive description (\ref{rec}) assumes
that, in particular, the inspector knows whether
the inspectee chose legal action or violation even after a
time period where the inspector did not inspect.
In practice, it may be rather questionable how the inspector
would obtain this knowledge.

In the games studied by Dresher \cite{Dr1962}, it actually does not
matter whether the inspector has this knowledge or not.
In Dresher's first game, the inspectee has only a single
intended violation, corresponding to $k=1$ in
our model (and, throughout, $r_i=1$ for all~$i$\,).
Then the lower-right entry in (\ref{rec}) given by
$v(n-1,m,0)-r_1$ is equal to $-1$.
In that case, because the inspectee has successfully
violated once and will not violate further, the game is
effectively over because the inspectee acts legally from
then on and will not be caught.
Then any action of the inspector is optimal, and so the
inspector can act as if the violation is still to take
place.
That is, if the inspector does know whether he is in
the game ${\Gamma(n-1,m,1)}$ or $\Gamma(n-1,m,0)$ (the latter
with added payoff $-1$ due to the uncaught violation), then
he can always act as if he is in the game $\Gamma(n-1,m,1)$
because that is the only situation where his strategy
matters.
Therefore, the recursive description is justified.

In the second game described by Dresher \cite{Dr1962}, the
inspectee tries to violate as often as possible.
This corresponds to our game $\Gamma(n,m,n-m)$ because the
inspectee can only violate once per time period and will
therefore not violate more than $n-m$ times because
otherwise he would be caught with certainty.
Then the bottom-left and bottom-right entries in (\ref{rec})
are $v(n-1,m,n-m)$ and $v(n-1,m,n-m-1)-1$, respectively.
However, the lower-left game $\Gamma(n-1,m,n-m)$ (where the
inspectee has ``missed out'' to violate during an
uninspected time period) is equivalent to the game
$\Gamma(n-1,m,n-m-1)$, that is, again a game with a maximal
number of intended violations.
The bottom-right game is the same, except for the added
$-1$ to the inspector's payoff, so again it does not matter
whether the inspector knows if the inspectee violated or
not.

Dresher \cite{Dr1962} gave explicit values for these two games as
in (\ref{k1nm}).
However, Dresher did not compute the optimal inspection
probabilities, because he would otherwise most likely have
noted that they are the same in the two games
$\Gamma(n,m,1)$ and $\Gamma(n,m,n-m)$.
These inspection probabilities are given by (\ref{p}).
A key aspect of our model is that they hold, independently
of~$k$, in the game $\Gamma(n,m,k)$ with the number $k$ of
intended violations as a new parameter.

Because of this independence of $k$, the equilibrium
strategy of the inspector, and the game value $v(n,m,k)$,
apply also to the game $\Gammaprime(n,m,k)$ without full
information where the inspector does \emph{not} know if a
violation occurred or not in an uninspected time period.
Namely, by induction the inspection strategy is the same in
the two games ${\Gammaprime(n-1,m,k)}$ and
${\Gammaprime(n-1,m,k-1)}$ which correspond to the two bottom
cells in (\ref{rec}), the latter with an additional loss of
$-r_k$ to the inspector, as long as the inspectee has still
an incentive to violate; if that is not the case, as in the
game $\Gammaprime(n-1,m,0)$ which has value zero, then any
inspection strategy is optimal and so the inspector should
act as if there are still violations to take place because
only then his action matters, as in Dresher's first game.

Formally, the game $\Gammaprime(n,m,k)$ without full
information is not described recursively.
However, it can be modelled as an extensive form game with
information sets \cite{Ku1953} that represent the
inspector's lack of information.
If we then change the game to the game with full
information, then these information sets are ``cut'', which
transforms $\Gammaprime(n,m,k)$ into the recursively
described game $\Gamma(n,m,k)$ in (\ref{rec}).
Because the inspector's behavior strategy in
$\Gamma(n,m,k)$ is the same at all information sets
obtained from ``cutting'' an information set $h$, say, in
the original game $\Gammaprime(n,m,k)$, it can also be defined
uniquely as the behavior at~$h$ and thus defines a behavior
strategy for $\Gammaprime(n,m,k)$.
In particular, the value of $\Gammaprime(n,m,k)$ stays the same 
at $v(n,m,k)$.
This (straightforward) manipulation of information sets is
described in detail in \cite{vS1991}.
In summary:

\begin{corollary}
\label{c-inf}
The equilibrium payoff and the equilibrium strategies for
the inspection game with full information described in
Theorem~\ref{t-main} also apply in the game
$\Gammaprime(n,m,k)$ without full information where the
inspector is not informed about the action of the inspectee
after a time period without inspection.
\end{corollary}

In the game $\Gammaprime(n,m,k)$ without full 
information, the inspectee has typically additional
equilibrium behavior strategies compared to $\Gamma(n,m,k)$.
As an example, let $n,m,k=3,1,2$ and $b=r_2=r_1=1$.
Then the bottom cells of (\ref{rec}) both correspond to the
game $\Gamma(2,1,1)$, with added payoff $-1$ in the
bottom-right cell.
In $\Gamma(2,1,1)$, which is (\ref{ABCD}) with
$A,B,C,D=-1,1,0,1$, the optimal strategies are $p=q=1/3$,
with $v(2,1,1)=-1/3$.
At the first stage in $\Gamma(3,1,2)$, they are $p=1/4$ and
$q=5/12$, which also applies to $\Gammaprime(3,1,2)$.
However, in the game $\Gammaprime(3,1,2)$, the inspector does
not know if the inspectee violated in the first time period
or not, which gives the inspectee additional optimal
behavior strategies for the second time period.
For example, the inspectee can violate with probability
$4/7$ if he acted legally in the first period and violate
with probability zero if he violated in the first period.
Another such coordinated different behavior in the second
time period would be to violate with probability zero
following legal action in the first period and to violate
with probability $4/5$ following a violation in the first
period.

We next discuss the rewards to the inspectee
$r_k,\ldots,r_1$ for successful violations, and the
corresponding scaled penalty $-b\cdot r_k$ to the inspectee
in~(\ref{rec}).
These parameters are new compared to von Stengel \cite{vS1991}, 
who proved Theorem~\ref{t-main} with $r_i=1$ for
$k\ge i\ge1$.
With general nonnegative rewards $r_i$, it seems that one
can dispense with the parameter~$k$ and simply assume that
only the first $k$ rewards $r_i$ are nonzero if the 
inspectee intends only $k$ violations.
In one respect this is a different game than when the
inspectee will not carry out more than $k$ violations,
because when all rewards are zero, the inspectee can violate
and be caught without penalty, which just terminates the
game; one may argue that this is an acceptable game outcome that just
has to be interpreted appropriately.
The main reason for the parameter $k$ in the recursive
description of the game is to identify the next reward to
the inspectee after a successful violation when the game
continues in the bottom-right cell in (\ref{rec}).
The number of intended violations serves as a ``counter''
for the rewards, which we have therefore numbered in the
order $r_k,r_{k-1},\ldots,r_1$.
Such a counter is needed for the recursive description in
one way or another.

The payoff $b\cdot r_k$ to the inspector for a caught
violation may seem strange in the game $\Gammaprime(n,m,k)$
where the inspector is not informed about~$k$.
However, we think it is justifiable to make the ``stakes''
of a violation proportional to $r_k$ even if the inspector
does not know~$r_k$, because the inspectee knows what is at
stake.
We have chosen this payoff as $b\cdot r_k$ because otherwise
the optimal inspection strategy would not be independent
of~$k$ as it is according to the solution~(\ref{p}).
In fact, the next theorem states that the payoffs in
(\ref{rec}) are as general as possible so that this
independence holds.
For simplicity, we assume that the marginal gain $r_j$ to
the inspectee for the next of $j$ remaining violations is
always positive, and that the game has a circular preference
structure.

\begin{theorem}
\label{t-gen} 
Suppose that $n,m,k$ are the number of time periods,
inspections, and intended violations in a zero-sum
inspection game where the inspectee can violate at most once
per time period, where his overall payoff depends only on
(and is strictly increasing in) the total number of
successful violations, and whether he is ever caught (in
which case the game terminates) or not.
Consider this game with full information and value $v(n,m,k)$.
Then the most general form of this game
fulfills $(\ref{mn})$ and $(\ref{m0})$, and for
$n>m>0$ and $k>0$ is the game
\begin{equation}
\label{g1}
\renewcommand\arraystretch{1.5}
\begin{array}{r|c|c|}
\cline{2-3}
p~~& v(n-1,m-1,k)& f(k) \\
\cline{2-3}
1-p~~& v(n-1,m,k) & v(n-1,m,k-1)-r_k\\ 
\cline{2-3}
\end{array}
\end{equation}
similar to $(\ref{rec})$, where we assume that $(\ref{g1})$
has a unique completely mixed equilibrium.
Here $f(k)$ is the marginal penalty and $r_k$ is the
marginal gain to the inspectee for the first of $k$
remaining violations, $r_k>0$\,.
Then the probability $p$ of inspection is independent of $k$
(so that it can be applied to the game without full
information) if and only if there is some $b>-1$ so that
$f(k)=b\cdot r_k$ for all~$k$, as in $(\ref{rec})$.
\end{theorem}

\proof{Proof.}
Consider the game with full information.
At the beginning of the game, we can assume $k\le n-m$
because the inspectee will not perform more than $n-m$
violations because he would otherwise be caught with
certainty.
For $1\le i\le n$, let $r_{k-i+1}$ be the marginal gain to
the inspectee for the $i$th successful violation, which by
assumption is strictly positive. 
Suppose that over the $n$ time periods, the inspectee
performs $i$ successful violations, $0\le i\le k$, and thus
gains $r_k+r_{k-1}+\cdots+r_{k-i+1}$~.
This is his payoff (and loss to the inspector), in
completely general form, if he is not caught.
If the inspectee is caught when attempting the $(i+1)$st
violation, then he pays the penalty $f(k-i)$, which is
subtracted from this sum (this penalty may include, for
example, repaying all previous gains); 
the inspector's payoff is then
$-r_k-r_{k-1}-\cdots-r_{k-i+1}+f(k-i)$.
Then $v(n,m,k)=0$ when $k=0$ or $m\ge n$ as in
(\ref{mn}) (for legal action throughout), and $v(n,0,k)$
given by (\ref{m0}).
For $n>m>0$ and $k>0$, the game with value $v(n,m,k)$ is
given by~(\ref{g1}), which is therefore the general form of
an inspection game under the stated assumptions.

If $k=1$, there is only one parameter $f(1)$ so
that we can set $b=f(1)/r_1$ and the inspector's strategy
can be applied to the game without full information; this is
essentially the first game by Dresher \cite{Dr1962}.
Hence, we can assume $k\ge2$.

Let $j\ge 2$ and suppose that the inspectee has performed
$k-j$ successful violations (and therefore, so far, gained
$r_k+r_{k-1}+\cdots+r_{j+1}$), that the inspector has
performed $m-1$ inspections, and that $n-3$ time periods
have passed.
The successful violations and the inspections have to take
place in different time periods, which is possible because
$k-j+m-1\le n-m-2+m-1=n-3$, and this occurs with positive
probability because of the mixed equilibrium at every stage
of the game.
Then at this stage there are three time periods, one
inspection, and $j$ intended violations remaining, and the
remaining game has value $v(3,1,j)$ and is of the form
\begin{equation}
\label{g2}
\renewcommand\arraystretch{1.5}
\begin{array}{r|c|c|}
\cline{2-3}
& v(2,0,j)& f(j) \\
\cline{2-3}
~& v(2,1,j) & v(2,1,j-1)-r_j\\ 
\cline{2-3}
\end{array}~.
\end{equation}
In the bottom left cell of (\ref{g2}), $v(2,1,j)$ is the value of 
\begin{equation}
\label{g3}
\renewcommand\arraystretch{1.5}
\begin{array}{r|c|c|}
\cline{2-3}
& v(1,0,j)& f(j) \\
\cline{2-3}
& v(1,1,j) & v(1,1,j-1)-r_j\\ 
\cline{2-3}
\end{array} ~,
\quad
\hbox{that is, of}
\quad
\begin{array}{r|c|c|}
\cline{2-3}
p~~& -r_j& f(j) \\
\cline{2-3}
1-p~~& 0 & -r_j\\ 
\cline{2-3}
\end{array}  ~.
\end{equation}
In the bottom right cell of (\ref{g2}), the inspectee
collects a reward of~$r_j$, and $v(2,1,j-1)$ is the value of
the game 
\begin{equation}
\label{g4}
\renewcommand\arraystretch{1.5}
\begin{array}{r|c|c|}
\cline{2-3}
& v(1,0,j-1)& f(j-1) \\
\cline{2-3}
& v(1,1,j-1) & v(1,1,j-2)-r_{j-1}\\ 
\cline{2-3}
\end{array} ~,
\quad
\hbox{that is, of}
\quad
\begin{array}{r|c|c|}
\cline{2-3}
p~~& -r_{j-1}& f({j-1}) \\
\cline{2-3}
1-p~~& 0 & -r_{j-1}\\ 
\cline{2-3}
\end{array} ~.
\end{equation}
The two games in (\ref{g3}) and (\ref{g4}) correspond to the
two cells in the bottom row of (\ref{g2}) and both must have
the same probability $p$ of inspection if this is to be
applied to the game without full information.
That is, according to (\ref{pp}), 
\begin{equation}
\label{pg34}
\frac{1-p}{p}
~=~
\frac1p-1
~=~
\frac{f(j)+r_j}{r_j}
~=~
\frac{f(j)}{r_j}+1
~=~ 
\frac{f({j-1})+r_{j-1}}{r_{j-1}}
~=~ 
\frac{f({j-1})}{r_{j-1}}+1~.
\end{equation}
For $j=2$, this shows $f(2)/r_2=f(1)/r_1=:b$, where $b>-1$
because $1/p-1>0$.
For $j=3$ it shows $f(3)/r_3=f(2)/r_2$, and so on,
so that $f(j)/r_j=b$ for all $1\le j\le k$, as claimed.
\endproof

Another question is if there is an intuitive reason that the
inspector's optimal strategy in $\Gamma(n,m,k)$ does not
depend on~$k$ (for any~$m$, not just for $m=1$ as in the
proof of Theorem~\ref{t-gen}).
For example, Ferguson and Melolidakis \cite{FeMe2000} have applied a
``game with finite resources'' due to Gale \cite{Ga1957} to a
different inspection game where the solution also applies
when one of the players lacks information.
However, we have not been able to apply the highly
symmetrical strategy in this game to our setting.
At present, the very canonical proof (see equation
(\ref{eq}) and onwards) of the explicit representation
(\ref{p}) and (\ref{v}) seems to be the best explanation.


To conclude this section, we discuss the solution of the
game $\Gamma(n,m,k)$ for some simple special cases.
If $m=1$, then it is easy to see that the inspector uses his
single inspection in the first $n-1$ time periods with
equal probability, which for the last time period is
multiplied with $1+b$, so if $b>0$ then higher probability
is given to the last period.

The case $b=0$, where a caught violation terminates the game
but no further penalty applies, has also some easily
described properties.
Then $s(n,m)={n\choose m}$ and thus $p=m/n$ in (\ref{p}),
which means that all $m$-sets of the $n$ time periods are
equally likely to be inspected.
Moreover, if $r_i=1$ for $k\ge i\ge1$, then
$t(n,m,k)=\sum_{i=1}^k{n-k\choose m}$ by (\ref{t}),
and $-v(n,m,k)=t(n,m,k)/s(n,m)$ can be interpreted as the
expected number of successful violations, as follows:
The inspectee is indifferent between all possible time
periods for choosing his $k$ violations, and thus gets payoff
$-v(n,m,k)$ if he violates in the first $k$ time periods.
Then, if the $m$-set of inspections does not include period~1
(with ${n-1\choose m}$ out of ${n\choose m}$ choices), the
first violation succeeds.
If this set also does not include period~2 (with further
${n-2\choose m}$ choices), then the first and second
violation succeed, and so on.

The probability $q$ of violation in the first period
depends on~$k$, and for $b=0$ and $r_i=1$ for $k\ge i\ge1$
has the following form.
If $k=1$, then $q=1/n$, independently of~$m$.
If $k=n-m$ (where the inspectee violates as often as
possible), then $q=1/(m+1)$, independently of~$n$.
Unfortunately, there is no straightforward simple extension
of these values for intermediate values of~$k$.
In general, we have only found complicated expressions for
the inspectee's strategy, which is why we have left it in
the form (\ref{q}) derived from the well-known
representation (\ref{qq}) in terms of the game payoffs.

\section{Non-zero-sum payoffs}
\label{s-nonzero}

In this section, we extend the zero-sum-game $\Gamma(n,m,k)$
to a non-zero-sum game $\hat\Gamma(n,m,k)$.
The reason to consider non-zero-payoffs is that a caught
violation as the outcome of the game is typically less
preferred by both inspector and inspectee than legal action,
because for the inspector it means the failure of the
inspection regime.
This is a standard assumption in inspection games, first
proposed by Maschler \cite{Ma1966}.

We denote the equilibrium payoffs in
$\hat\Gamma(n,m,k)$ by $v(n,m,k)$ for the inspector and
by $w(n,m,k)$ for the inspectee (which are unique
as shown in Theorem~\ref{t-unique} below).
The reference payoff for legal action throughout is zero for
both players.
As before, we assume that the inspectee acts legally if the
inspector can inspect in every remaining period, that is, 
\begin{equation}
\label{nonmn}
v(n,m,k)=w(n,m,k)=0\qquad
\hbox{if }m\ge n.
\end{equation}
Also as before, if the inspector has run out of inspections,
then the inspectee collects a nonnegative reward $r_k,
r_{k-1},\ldots $ which is a cost to the inspector for each
remaining period up to the maximum number $k$ of intended 
violations, that is,
\begin{equation}
\label{nonm0}
-v(n,0,k)=
w(n,0,k)=
\sum_{i=1}^{\min\{k,n\}}
r_{k+1-i}~.
\end{equation}
For the case that a violation is caught we introduce two
parameters $a$ and $b$ as costs to inspector and inspectee
(scaled by the reward $r_k$ for a successful violation),
where 
\begin{equation}
\label{ab}
0<a<1,\qquad
b\ge0,
\end{equation}
so that for $n>m>0$ and $k>0$ the game $\hat\Gamma(n,m,k)$
(with full information) has the following recursive
description:
\begin{equation}
\label{nonzero}
\setlength{\unitlength}{1.2em}
\lower4.8\unitlength\hbox{%
\begin{picture}(27,11)(0,-0.5)
\put(0,10){\line(0,-1){10}}
\put(6,10){\line(0,-1){10}}
\put(25,10){\line(0,-1){10}}
\put(15,10){\line(0,-1){10}}
\put(0,8){\line(1,0){25}}
\put(0,10){\line(3,-1){6}}
\put(0,10){\line(1,0){25}}
\put(0,4){\line(1,0){25}}
\put(0,0){\line(1,0){25}}
\put(1.9,8.5){\mm{{inspector}}}
\put(4.1,9.5){\mm{{inspectee}}}
\put(3,6){\mm{{inspection}}}
\put(9.0,7){\lm{$w(n{-}1,{m{-}1},k)$}}
\put(6.3,5){\lm{$v(n{-}1,{m{-}1},k)$}}
\put(3,2){\mm{{no inspection}}}
\put(9.0,3){\lm{$w(n{-}1,m,k)$}}
\put(6.3,1){\lm{$v(n{-}1,m,k)$}}
\put(21.0,3){\mm{$w(n{-}1,m,{k{-}1})+{r_k}$}}
\put(18.9,1){\mm{$v(n{-}1,m,{k{-}1})-{r_k}$}}
\put(23,7){\mm{$-b\cdot{r_k}$}}
\put(17,5){\mm{${-a\cdot{r_k}}$}}
\put(10.5,9){\mm{{legal action}}}
\put(20,9){\mm{{violation}}}
\put(5.5,4){\mm{{$\downarrow$}}}
\put(25.5,4){\mm{{$\uparrow$}}}
\put(15,8.5){\mm{{$\leftarrow$}}}
\put(15,-.5){\mm{{$\rightarrow$}}}
\end{picture}}
\end{equation}
In (\ref{nonzero}), the arrows represent the circular
preferences of the players, which have been proved for
$\Gamma(n,m,k)$ as~(\ref{ineq}).
In particular, if $k=1$, then the bottom-right cell in
(\ref{nonzero}) for an uncaught violation has payoff $-r_k$
to the inspector, whereas the top-right cell has payoff
$-a\cdot r_k$.
Because $a<1$, the inspector therefore prefers a caught
violation to an uncaught one, as it should be the case.

Due to (\ref{ab}), the game $\hat\Gamma(n,m,k)$ does not
include the zero-sum game $\Gamma(n,m,k)$ as a special case.
However, the more general conditions $a<1$ and $b>-1$ do
include it when $a=-b$.
The following theorem is essentially a corollary to
Theorem~\ref{t-main}.

\begin{theorem}
\label{t-non} 
Let $n,m,k$ be nonnegative integers,
let the reals $a$ and $b$
be as in~$(\ref{ab})$,
and let $r_k,r_{k-1},\cdots,r_1\ge 0$.
Define $s(n,m,k)$
as in~$(\ref{snm})$,
$t(n,m,k)$ as in~$(\ref{t})$,
and $\hat s(n,m)$ by
\begin{equation}
\label{hatsnm}
\hat s(n,m)=\sum_{i=0}^m{n\choose i}(-a)^{m-i}~.
\end{equation}
Then the non-zero-sum game $\hat\Gamma(n,m,k)$ defined by
$(\ref{nonzero})$ for $n>m>0$ and $k>0$ and by $(\ref{nonmn})$ and
$(\ref{nonm0})$ otherwise has equilibrium payoffs to
inspector and inspectee
\begin{equation}
\label{vw}
v(n,m,k)=\frac{-t(n,m,k)}{\hat s(n,m)}~,
\qquad
w(n,m,k)=\frac{t(n,m,k)}{s(n,m)}~.
\end{equation}
For $n>m>0$ and $k>0$, the game $(\ref{nonzero})$ has a
completely mixed equilibrium where the inspector inspects
with probability $p$ according to $(\ref{p})$, and the
inspectee violates with probability $q$ according to
$(\ref{q})$. 
This is the unique subgame perfect Nash equilibrium (SPNE) of the game,
unless $r_{k+1-i}=0$ for $1\le i\le\min\{k,n-m\}$, in which
case all entries in $(\ref{nonzero})$ are zero and the
players can play arbitrarily.  
Each player's strategy is the min-max strategy for the
payoffs of his opponent.
\end{theorem}

\proof{Proof.}
If we modify the game $\hat\Gamma(n,m,k)$ to a zero-sum game
with the payoffs $v(n,m,k)$ to the inspector (and thus
$-v(n,m,k)$ to the inspectee), then it fulfills the
assumptions of Theorem~\ref{t-main} with $b=-a>-1$.
In this game, the inspector prefers not to inspect when the
inspectee acts legally and to inspect when the inspectee
violates, as shown with the vertical arrows in~(\ref{nonzero}).
The inspectee's strategy is as in (\ref{q}) and
is a min-max strategy for the inspector's payoff.
It makes the inspector indifferent between his two actions,
with the inspector's payoff $v(n,m,k)$ as in (\ref{vw}).
Note that in (\ref{hatsnm}), $\hat s(n,m)>0$ for
$0\le m\le n$ due to the alternative representation
(\ref{salt}) where $1+b=1-a>0$.

Similarly, if we modify the game $\hat\Gamma(n,m,k)$ to a
zero-sum game based on the payoffs $w(n,m,k)$ to the
inspectee (and thus $-w(n,m,k)$ to the inspector), then it
fulfills also the assumptions of Theorem~\ref{t-main} with
$b\ge0>-1$.
Then the inspectee prefer to act legally when he is
inspected and to violate otherwise, as shown with the
horizontal arrows in~(\ref{nonzero}).
In this game, the inspector's strategy is given by
(\ref{p}), and is a min-max strategy for the inspectee's
payoff.
It makes the inspectee indifferent between his two
actions in (\ref{nonzero}), with the inspectee's payoff
$w(u,m,k)$ as in~(\ref{vw}).

So the game in (\ref{nonzero}) has a circular preference
structure and a unique mixed equilibrium as described
(except when $r_{k+1-i}=0$ for $1\le i\le\min\{k,n-m\}$),
which by induction defines the unique SPNE of
$\hat\Gamma(n,m,k)$.  
\endproof

In Theorem~\ref{t-non}, the inspector's strategy in
$\hat\Gamma(n,m,k)$ does not depend on~$k$.
As argued in \S\ref{s-strategy}, this strategy can
therefore also be applied to the game $\hat\Gammaprime(n,m,k)$
without full information.
That is, we obtain the analogous statement to Corollary~\ref{c-inf}.

\begin{corollary}
\label{c-nonzinf}
The equilibrium payoff and the equilibrium strategies for
the non-zero-sum inspection game described in
Theorem~\ref{t-non} with full information also apply in the
game $\hat\Gammaprime(n,m,k)$ without full information.
\end{corollary}

As mentioned after Corollary~\ref{c-inf}, the game
$\Gammaprime(n,m,k)$ without full information may
have additional equilibrium strategies for the inspectee,
which applies in the same way to the game
$\hat\Gammaprime(n,m,k)$.

Because the games $\hat\Gamma(n,m,k)$ and
$\hat\Gammaprime(n,m,k)$ are not zero-sum, the
question arises if they have other Nash equilibrium payoffs.
The following theorem asserts that this is not the case.

\begin{theorem}
\label{t-unique}
All Nash equilibria of the non-zero-sum inspection game
$\hat\Gamma(n,m,k)$ and the game $\hat\Gammaprime(n,m,k)$
without full information have the payoffs described in
Theorem~\ref{t-non}.  
\end{theorem}

\proof {Proof.}
Consider first the game $\hat\Gamma(n,m,k)$ with full
information.
Let the game be represented as an extensive game.
Call a \emph{stage} of the game a particular time period
together with the history of past actions.
At each stage, we let, as in \cite{vS1991}, the
inspector move first and the inspectee second, where the
decision nodes of the inspectee belong to a two-node
information set so that the inspectee is not informed about
the action of the inspector at the current stage, but knows
everything else.
The information set of the inspector is a singleton
(this is different in the game $\hat\Gammaprime(n,m,k)$ that
we consider later).

Consider a Nash equilibrium of this game.
Suppose that there is a stage of the game that is reached
with positive probability where the players do not behave
according to the SPNE described in Theorem~\ref{t-non}, and
let there be no later such stage.
That is, each of four cells in (\ref{nonzero}) at this stage
either has the SPNE payoffs as entries or is reached with
probability zero.
We claim that the equilibrium property is violated at this
stage.
If all cells have positive probability, then at least one
player gains because they do not play the unique equilibrium
at this stage.
If some cells have probability zero, then one player plays
deterministically. 
For example, suppose the inspectee acts legally.
Then if the inspector inspects with positive
probability, he gets the SPNE payoff $v(n-1,m-1,k)$.
However, this is not his best response, because when he
does not inspect at this stage, he gets at least
$v(n-1,m,k)$ because that is also his min-max payoff which
he can guarantee by playing a max-min strategy after no
inspection at the current stage.
Because we are in equilibrium, the inspector therefore
responds with no inspection to the inspectee's certain legal
action at this stage.
However, then the inspectee can improve on his SPNE payoff
$w(n-1,m,k)$ by violating and subsequently playing a max-min
strategy, which contradicts the assumed equilibrium.
This reasoning follows from the strictly circular payoff
structure in (\ref{nonzero}) and holds for any assumed
unplayed strategy.
Hence, players always mix and the SPNE of the game
$\hat\Gamma(n,m,k)$ is its unique Nash equilibrium (in
behavior strategies, as always).

The crucial condition used is that SPNE payoffs are min-max
payoffs, and the argument is similar to an analogous known
result on finitely repeated games where all stage
equilibrium payoffs are min-max payoffs (see Osborne and
Rubinstein, \cite[Proposition~155.1]{OsRu1994}).

Before we consider the game $\hat\Gammaprime(n,m,k)$, we
discuss a potential ``threat'' of the inspectee to use a
violation even in the case $m\ge n$ when the inspector can
inspect in every remaining period.
If we assume that in this case the inspector has the choice
not to inspect, this defines 
a game where legal action gives payoff zero to both players
but violation gives a negative payoff to both players.
The min-max payoff to the inspector is then $-a\cdot r_k$
rather than zero, when the inspectee irrationally violates
later (which is his ``threat'') and the inspector inspects.
If we use this payoff $-a\cdot r_k$ in the bottom-left cell
of (\ref{nonzero}) rather than the assumed zero SPNE payoff
$v(n-1,n-1,k)$, we show that nevertheless 
$-a\cdot r_k>v(n-1,n-2,k)$ so that the preceding reasoning
still applies, that is, the inspector's best response to
legal action at the current stage (and violation later) is
still no inspection.
Namely, by (\ref{vw}), and (\ref{s0}) and (\ref{sb})  with
$b=-a$ and $m=n-1$, 
\[
v(n-1,n-2,k)=\frac{-t(n-1,n-2,k)}{\hat s(n-1,n-2)}=
\frac{-r_k{n-2\choose n-2}}
{((1-a)^{n-1}-{n-1\choose n-1})/(-a)}
=\frac{-a\cdot r_k}{1-(1-a)^{n-1}}
\]
and thus
$-a\cdot r_k>v(n-1,n-2,k)$ as claimed because $0<a<1$ by
(\ref{ab}).
That is, the inspector still prefers not to inspect in
response to legal action and a later ``threatened''
violation that will be caught.
Hence there is no ``threat'' of the inspectee that could
induce a Nash equilibrium other than the SPNE.

Consider now a Nash equilibrium in behavior strategies of
the game $\hat\Gammaprime(n,m,k)$ without full information.
The inspector's lack of information is represented by
information sets of the inspector that comprise multiple
decision nodes with the same history of the inspector's own
past actions, but different past actions of the inspectee at
the stages where the inspector did not inspect.
Consider such an information set $h$ of the inspector that is
reached with positive probability where the inspector does
\emph{not} use the min-max strategy against the inspectee
in Theorem~\ref{t-non}, and assume that there is no later
information set of this kind.
Then at this stage, that is, for \emph{all} information sets
of the inspectee that immediately follow this move of the
inspector at $h$, the inspectee will have the same action
(legal action or violation) as a best response, which he
therefore chooses with certainty because we are in
equilibrium.
However, in response the inspector would have to make a move
at~$h$ against which the moves of the inspectee are not
optimal.
This contradicts the equilibrium property.
Hence, the inspector has to choose the min-max strategy
throughout, so that the inspectee's payoff is as in 
Theorem~\ref{t-non}.

Now suppose that the inspector's payoff is different from
his min-max payoff.
This has to be a larger payoff because the inspector can
guarantee his min-max payoff by playing a max-min strategy.
Then at some information set $h$ of the inspector that is
reached with positive probability, again looking at the
latest such set, the inspectee does at this stage not play a
min-max strategy against the inspector, that is, a strategy
that does not equalize the inspector's payoffs.
To this the inspector plays a unique pure best response at~$h$.
This response is different from the inspector's strategy in
Theorem~\ref{t-non}, but we have just shown that this cannot
be the case.  

Hence, the players' payoffs are uniquely given according to
Theorem~\ref{t-non}, as claimed.
\endproof

\section{Inspector leadership}
\label{s-lead}

The game by Dresher \cite{Dr1962} with a single intended violation
has been studied by Maschler \cite{Ma1966} in a \emph{leadership}
variant where the inspector can announce and commit to his
mixed strategy.
We extend these considerations to our game with $k$ intended
violations, and simplify some of Maschler's arguments.

A two-player game in strategic form is changed to a
leadership game by declaring one player as leader and the
other as follower.
The leader chooses and commits to a strategy about which the
follower is fully informed and chooses, as in a subgame
perfect equilibrium, a best response to every commitment of
the leader.
Both players receive the payoffs of the original game.
A leadership game is often called a ``Stackelberg game'',
following von Stackelberg \cite{vonSta1934} who modified in this manner
the simultaneous model of quantity competition by Cournot
\cite{Co1838} to a sequential game.
We consider the leadership game for the mixed extension of a
finite two-player where the leader can commit to a mixed
strategy, as analyzed in full generality by von Stengel and
Zamir \cite{vSZa2010}.

Inspection games model situations where an inspectee is
obliged to act legally and hence cannot openly declare that
he will violate.
However, the inspector can become a leader and commit to a
mixed strategy, using a ``roulette wheel'' or other
randomization device that decides with a verifiable
probability in each time period (simultaneously to the
choice of the inspectee) whether to inspect or not.
Maschler \cite{Ma1966} observed that in a non-zero-sum recursive
game, similar to (\ref{nonzero}) for $k=1$, the inspector
can commit to essentially the
same mixed strategy as before, but that the inspectee acts
legally with certainty as long as inspections remain.
We first consider a general $2\times2$ game as it arises in
our context.

\begin{figure}[bht]
\[\includegraphics[height=70mm]{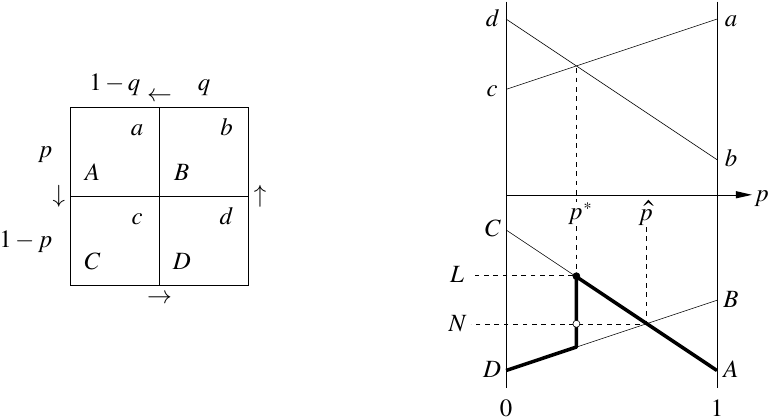}\]
\caption{Left: $2\times 2$ game with probability $p$ for playing the
top row and $q$ for playing the right column.
Right: Payoffs to column and row player if 
(\ref{zineq}) and (\ref{leftbetter}) hold in the leadership
game where the row player commits to~$p$.}
{}
\end{figure} 

\begin{proposition}
\label{p-22commit}
Consider the $2\times 2$ game on the left in 
Fig.~1
where the payoffs $A,B,C,D$ to the row player and $a,b,c,d$
to the column player fulfill 
\begin{equation}
\label{zineq}
A<C, \quad B>D,\quad
a>b, \quad c<d~.
\end{equation} 
Let
\begin{equation}
\label{pstar}
p^*=\frac{d-c}{a-b+d-c}
\end{equation} 
and assume that
\begin{equation}
\label{leftbetter}
p^*A+(1-p^*)C> p^*B+(1-p^*)D~.
\end{equation} 
Then the game has a unique mixed equilibrium where $p^*$ is
the equilibrium probability that the row player plays the
top row,
with Nash payoff $N=\frac{BC-AD}{B-D+C-A}$ to the row player.
In the leadership game where the row player is the leader
and can commit to a mixed strategy $p$, the unique subgame
perfect equilibrium is that the row player commits to $p^*$
and the column player responds with $q=1$ if $p<p^*$ and
$q=0$ if $p\ge p^*$, in particular with $q=0$ on the
equilibrium path where $p=p^*$.
In the leadership game, the payoff to the leader is
$L=p^* A+(1-p^*)C$, and $L>N$.
The payoff to the follower is $p^* a+(1-p^*)c$, the same as
in the simultaneous game.
\end{proposition}

\proof{Proof.}
By (\ref{zineq}), the game has a unique mixed equilibrium
where the row player plays $p^*$ and the column player plays
$q^*=\frac{C-A}{B-D+C-A}$, and the row player gets payoff
$N$ and the column player gets $p^*a+(1-p^*)c$.

The claims about the leadership game can be seen from
the right picture in Figure~1, which shows the players'
payoffs as a function of~$p$.
For illustration, the column player's payoffs are assumed to
be positive (as it typically holds in our inspection games,
with the exception of the payoff~$b$) and those of the
column player as negative, here for the case that $B>A$ (so
that (\ref{leftbetter}) can only hold if $C>D$).

For $p<p^*$ the follower's best response is the right column 
($q=1$), with expected payoff $pb+(1-p)d$ to the follower
and $pB+(1-p)D$ to the leader.
For $p>p^*$ the follower's best response is the left column
($q=0$), with expected payoff $pa+(1-p)c$ to the follower
and $pA+(1-p)C$ to the leader.
For the commitment to $p=p^*$ the follower is indifferent
and in principle could reply with any $q$ in $[0,1]$.
The payoff to the leader as a function of $p$ is shown as
the bold line in the figure, including the vertical part for
$p=p^*$ for $q\in[0,1]$.
By (\ref{zineq}), this leader payoff is increasing in $p$
for $p<p^*$ and decreasing in $p$ for $p>p^*$, so it has its
maximum $L$ if $p=p^*$ and, by (\ref{leftbetter}), if the
follower's response is $q=0$, shown by a full dot in the
picture.
This reaction of the follower defines in fact the unique
SPNE in the leadership game, because for $p=p^*$ the
follower, even though indifferent, has to choose the
response, here $q=0$, that maximizes the leader's payoff,
because otherwise the leader could \emph{induce} this
behavior by changing his commitment to $p^*+\varepsilon$ for
some arbitrarily small positive $\varepsilon$, which
contradicts the SPNE condition.

This has essentially been observed by Maschler \cite{Ma1966}, who
postulated a ``Pareto-optimal'' response of the follower if
he is indifferent, and noted that otherwise the leader
can get a payoff arbitrarily close to $L$ with a commitment
to $p^* +\varepsilon$.  
The SPNE argument has been made by Avenhaus, Okada, and
Zamir \cite{AvOkZa1991}, and in generality by von Stengel and Zamir
\cite{vSZa2010} who also give further references.  

In addition to the leader payoff $L$, Fig.~1
shows the Nash payoff $N$ (with a hollow dot) further below
on the vertical line, which is less than $L$ because it is
given by $N=q^*L+(1-q^*)M$ where $M=p^*B+(1-p^*)D$ is the
minimum payoff to the leader if the follower responds to
$p^*$ by choosing the right column $q=1$, and $L>M$ by
(\ref{leftbetter}).
If $B>A$ and thus $C>D$, then $N$ is also the max-min payoff
to the row player where his expected payoffs are the
same for both columns, when he plays his max-min strategy 
$\hat p=\frac{C-D}{B-A+C-D}$, also shown in the picture.
\endproof

We want to apply Proposition~\ref{p-22commit} to the
inspection game (\ref{nonzero}).
Similar to Maschler \cite[p.~18]{Ma1966}, the leadership game for
$n$ time periods, $m$ inspections, and $k$ intended
violations is described as follows: 
\begin{equation}
\label{commit}
\setlength{\unitlength}{1.2em}
\lower4.8\unitlength\hbox{%
\begin{picture}(27,11)(0,-0.5)
\put(0,10){\line(0,-1){10}}
\put(6,10){\line(0,-1){10}}
\put(25,10){\line(0,-1){10}}
\put(15,10){\line(0,-1){10}}
\put(0,8){\line(1,0){25}}
\put(0,10){\line(3,-1){6}}
\put(0,10){\line(1,0){25}}
\put(0,4){\line(1,0){25}}
\put(0,0){\line(1,0){25}}
\put(1.9,8.5){\mm{{inspector}}}
\put(4.1,9.5){\mm{{inspectee}}}
\put(3,7){\mm{{roulette wheel}}}
\put(3,6){\mm{{calls for}}}
\put(3,5){\mm{{inspection}}}
\put(9.0,7){\lm{$w(n{-}1,{m{-}1},k)$}}
\put(6.3,5){\lm{$u(n{-}1,{m{-}1},k)$}}
\put(3,3){\mm{{roulette wheel}}}
\put(3,2){\mm{{calls for}}}
\put(3,1){\mm{{no inspection}}}
\put(9.0,3){\lm{$w(n{-}1,m,k)$}}
\put(6.3,1){\lm{$u(n{-}1,m,k)$}}
\put(21.0,3){\mm{$w(n{-}1,m,{k{-}1})+{r_k}$}}
\put(18.9,1){\mm{$u(n{-}1,m,{k{-}1})-{r_k}$}}
\put(23,7){\mm{$-b\cdot{r_k}$}}
\put(17,5){\mm{${-c(k)}$}}
\put(10.5,9){\mm{{legal action}}}
\put(20,9){\mm{{violation}}}
\end{picture}}
\end{equation}
If the assumptions of Proposition~\ref{p-22commit} are met,
then in this leadership game the inspector chooses the same
strategy as in the simultaneous game so that the inspectee
is indifferent between legal action and violation.
However, the inspectee acts legally as long as $m>0$, that
is, there will never be a caught violation.
For that reason, the result will hold for any negative cost
$-c(k)$ to the inspector in that cell of the table.

In the game (\ref{commit}), the inspectee as follower
should, by Proposition~\ref{p-22commit},
get the same recursively defined payoff $w(n,m,k)$ as in
Theorem~\ref{t-non}, but the inspector gets a new payoff
$u(n,m,k)$. 
The following consideration shows what this payoff should
be.
First, if $m=0$, then the inspectee can and will safely use
his $k$ intended violations, as far as possible, in each of
the remaining $n$ time periods, so that as in (\ref{nonm0}),
\begin{equation}
\label{comm0}
-u(n,0,k)=
w(n,0,k)=
\sum_{i=1}^{\min\{k,n\}}
r_{k+1-i}~,
\end{equation}
as well as
\begin{equation}
\label{commn}
u(n,m,k)=w(n,m,k)=0\qquad
\hbox{if }m\ge n.
\end{equation}
For $n>m>0$ and $k>0$, the game (\ref{commit}) applies, where
the inspectee gets the same payoff $w(n,m,k)$ for legal
action and violation, given by (\ref{vw}).
In particular, this is the inspectee's payoff if he always
acts legally, as we assume he does in the leadership game.
Once the inspector has run out of inspections, the inspectee
gets the same payoff $w(n,0,k)$ as in (\ref{nonm0}) and
(\ref{comm0}), which is the negative of the inspector's payoff.
By induction, the inspector's payoff should therefore in
general simply be $u(n,m,k)=-w(n,m,k)$.

In the following theorem, subgame perfection refers to the
leadership game that assumes best responses of the follower
even off the equilibrium path, namely for all other
inspection probabilities that the inspector could commit~to.
In terms of information about the history of the game, the
probability of the ``roulette wheel'' at each stage is a
function of $n$ and $m$ but not of~$k$, as before.

\begin{theorem}
\label{t-com} 
Let $n,m,k$ be nonnegative integers, 
$r_k,r_{k-1},\cdots,r_1\ge 0$, $b\ge0$ and $c(k)>0$.
Then in the leadership game defined by $(\ref{commn})$,
$(\ref{comm0})$, and $(\ref{commit})$ for $n>m>0$ and $k>0$,
the unique subgame perfect equilibrium payoff is $w(n,m,k)$
as in $(\ref{vw})$ to the inspectee, and
$u(n,m,k)=-w(n,m,k)$ to the inspector.
The inspector commits to the same inspection probability
$p^*=p$ as in the game with simultaneous actions in each
time period according to $(\ref{p})$.
For $m>0$, the inspectee always acts legally, and for
$m=0$ he violates in each remaining time period, up to $k$
times, as in $(\ref{comm0})$.  
Compared to the simultaneous game
in Theorem~\ref{t-non}, the inspector's cost is
smaller by the factor $\hat s(n,m)/s(n,m)$;
the inspectee's payoff is the same.  
\end{theorem}

\proof{Proof.}
By induction on $n$.
For $m=0$, we have $s(n,m)=s(n,0)=1$, so that 
$w(n,0,k)=t(n,0,k)$ which fulfills (\ref{comm0}) by (\ref{t}).
Similarly, $t(n,m,k)=0$ if $m\ge n$, which implies
(\ref{commn}).
In the same way, (\ref{comm0}) holds for the inspector's
payoff $u(n,0,k)$, and so does (\ref{commn}).

Let $n>m>0$ and $k>0$.
In (\ref{commit}), the probability for inspection $p$ should
make the inspectee indifferent between legal action and
violation, so that, as in Theorem~\ref{t-non}, the inspectee
gets the payoff $w(n,m,k)$ as claimed, and $p$ is given
by~(\ref{p}).
If the inspectee always acts legally, then the inspector's
payoff is recursively defined by 
\begin{equation}
\label{urec}
u(n,m,k)= p\cdot u(n-1,m-1,k)+(1-p)\cdot u(n-1,m,k)
\end{equation}
as in (\ref{eq}), which has been shown to be true in
(\ref{eqt1}) and (\ref{tt}).

It remains to show that (\ref{leftbetter}) in
Proposition~\ref{p-22commit} applies, that is, the inspectee
indeed acts legally because the inspector's payoff for legal
action is higher than for violation.
By (\ref{commit}), using (\ref{urec}), this is equivalent to
\begin{equation}
\label{legalbetter}
u(n,m,k)> p\cdot (-c(k))+(1-p)\cdot (u(n-1,m,k-1)-r_k).
\end{equation}
Now, because $u(n,m,k)=-w(n,m,k)$, we know that, analogous to
(\ref{eq2}) which has been shown with (\ref{eqt2}) and
(\ref{ttt}),
\[
u(n,m,k)=p\cdot b\cdot r_k+(1-p)\cdot(u(n-1,m,k-1)-r_k)\,,
\]
so that (\ref{legalbetter}) is equivalent to
$b\cdot r_k>-c(k)$, which is true.
So the recursive equation for $u(n,m,k)$ in (\ref{urec})
is indeed justified.

To compare the payoff $u(n,m,k)$ to the inspector in the
leadership game with his payoff $v(n,m,k)$ in the game with
simultaneous action in each time period, (\ref{vw}) gives
\begin{equation}
\label{compare}
u(n,m,k)
=\frac{-t(n,m,k)}{s(n,m)}
=\frac{\hat s(n,m)}{s(n,m)}\cdot
\frac{-t(n,m,k)}{\hat s(n,m)}
=\frac{\hat s(n,m)}{s(n,m)}\cdot
v(n,m,k)
\end{equation}
as claimed, where the factor $\hat s(n,m)/s(n,m)$ is smaller
than $1$ by (\ref{snm}) and (\ref{hatsnm}), possibly
significantly so, depending on the parameters $a$ and $b$ in
(\ref{ab}).  
\endproof

We conclude this section with two further observations.
The first is that even if one does not see that $u(n,m,k)$
is just $-w(n,m,k)$ as described following (\ref{commit}),
the recursive equation (\ref{urec}) shows what $u(n,m,k)$
should be, and also why $t(n,m,k)$ should be as in~(\ref{t}).
Namely, suppose that we do not yet know $t(n,m,k)$ and 
assume that $u(n,m,k)=-t(n,m,k)/s(n,m)$.
Then, by (\ref{p}), equation (\ref{urec}) is equivalent to
\[
\frac{-t(n,m,k)}{s(m,n)}
=
\frac{s(m-1,n-1)}{s(n,m)}
\cdot
\frac{-t(n-1,m-1,k)}{s(m-1,n-1)}
~+~
\frac{s(m,n-1)}{s(m,n)}
\cdot
\frac{-t(n-1,m,k)}{s(m,n-1)}
\]
which is just (\ref{eqt1}). 
So $t(n,m,k)$ fulfills the equation of a ``generalized
Pascal triangle''
\begin{equation}
\label{trecu}
t(n,m,k) =  t(n-1,m-1,k) + t(n-1,m,k)
\end{equation}
in $n$ and $m$ (with $k$ as a fixed parameter) and ``base
cases'' (by (\ref{comm0}) and because $s(n,0)=1$)
\[
t(n,0,k)=\sum_{i=1}^k r_{k+1-i}~,
\qquad t(n,n,k)=0~.
\]
Writing down the numbers $t(n,m,k)$ in a triangle as
functions of $r_k$, $r_{k-1}$, etc., one sees that the 
smallest $n$ where $r_{k+1-i}$ appears in $t(n,m,k)$
is for $n=i$ and $m=0$.
Due to (\ref{trecu}), this becomes the root $0\choose0$ of
an ordinary ``Pascal triangle'' for the coefficient of
$r_{k+1-i}$ in $t(n,m,k)$, which is therefore ${n-i\choose
m}$, as in~(\ref{t}). 

The second observation addresses the question if the
inspector always prefers the inspectee to act legally in the
game (\ref{nonzero}) where his payoffs are given by
$v(n,m,k)$, and not recursively by $u(n,m,k)$ as in
(\ref{commit}).
As an application of Proposition~\ref{p-22commit}, this
would apply to a leadership game where the inspector can
only commit to the probability of inspecting in the first
time period, but acts without commitment in all subsequent
periods.
This is not a very natural game to look at, but the
preference of the inspector in the simultaneous game is
nevertheless of interest.
As expected, the inspector indeed prefers that the inspectee
acts legally.
We found only a relatively long -- but ``canonical'' --
proof, which we present here for its possible interest
concerning the manipulation of sums of binomial
coefficients.

\begin{theorem}
\label{t-yeslegal}
Consider the game $(\ref{nonzero})$ with entries as in
Theorem~\ref{t-non} as a $2\times 2$ game on the left in
Fig.~1, and the equilibrium probability $p^*=p$ as in
$(\ref{p})$ for the inspector.
Then $(\ref{leftbetter})$ in Proposition~\ref{p-22commit}
applies, that is,
the inspector prefers that
in response to $p^*$
the inspectee acts legally.
\end{theorem}

\proof{Proof.}
In the game (\ref{nonzero}), we have $A,C,D$ as in
(\ref{ACD}) and $B=-a\cdot r_k$, where $-a>-1$ by
(\ref{ab}).
To show (\ref{leftbetter}) directly we would have
to compare terms involving $s(n,m)$, $t(n,m,k)$, and
$\hat s(n,m)$ according to (\ref{p}), (\ref{hatsnm}) and
(\ref{vw}).
Instead, we apply Theorem~\ref{t-main} 
with $b=-a$ to the zero-sum game (\ref{ABCD}) with entries
$A,B,C,D$, where the inspector has the max-min strategy of
inspecting with probability
\begin{equation*}
\hat p =\frac{\hat s(n-1,m-1)}{\hat s(n,m)}~.
\end{equation*}
Because $A<B$ and $C>D$ as shown in (\ref{ineq}),
$\hat p$ is also the probability that equalizes the expected
payoffs to the row player for the two columns of the game,
$\hat p=\frac{C-D}{B-A+C-D}$, as shown on the right in
Fig.~1.
Then (\ref{leftbetter}) holds if $\hat p > p^*$,
because the expected payoff for the left column
is $pA+(1-p)C=C+p(A-C)$ which is strictly decreasing in~$p$,
and for the right column it
is $pB+(1-p)D=D+p(B-D)$ which is strictly increasing in~$p$,
so that
$\hat p > p^*$
implies
\[
p^*A+(1-p^*)C
~>~
\hat pA+(1-\hat p)C
~=~
\hat pB+(1-\hat p)D
~>~ 
p^*B+(1-p^*)D
\]
that is, (\ref{leftbetter}).

To show $\hat p>p^*=p$, we define 
\[
S(n,m,x)=
\sum_{i=0}^m{n\choose i}(x-1)\ttp{i},
\]
so that by (\ref{hatsnm}) and (\ref{p})
\[
\hat s(n,m)=S(n,m,1-a),
\qquad
s(n,m)=S(n,m,1+b).
\]
As in (\ref{salt}) we have
\begin{equation}
\label{Sx}
S(n,m,x)=
\sum_{i=0}^m{n-1-i\choose m-i}x\ttp{i} .
\end{equation}
Then we want to show, for $n>m>0$, that
\begin{equation}
\label{phat}
\hat p =\frac{\hat s(n-1,m-1)}{\hat s(n,m)}
~>~ p =\frac{s(n-1,m-1)}{s(n,m)}~,
\end{equation}
or equivalently, by (\ref{ssum}),
\[
\frac{1-\hat p}{\hat p}
=\frac{\hat s(n-1,m)}{\hat s(n-1,m-1)}
~<~
\frac{1-p}{p}
=\frac{s(n-1,m)}{s(n-1,m-1)}~,
\]
that is,
\[
\frac{S(n-1,m,1-a)}{S(n-1,m-1,1-a)}
~<~
\frac{S(n-1,m,1+b)}{S(n-1,m-1,1+b)}~,
\]
which clearly holds if
${S(n-1,m,x)}/{S(n-1,m-1,x)}$ is strictly increasing in $x$
for $x>0$, which is what we will show.
By (\ref{sb}) we have with $x=1+b$ 
\[
S(n-1,m,x)=(x-1)\cdot S(n-1,m-1,x)+{n-1\choose m}
\]
so that 
\[
\frac{S(n-1,m,x)}{S(n-1,m-1,x)}
= x-1~+~\frac{{n-1\choose m}}{S(n-1,m-1,x)}
\]
where we want to show that this term has positive derivative
with respect to $x$, that is,
\[
\frac d{dx}\left(\frac{S(n-1,m,x)}{S(n-1,m-1,x)}\right)
~=~
1-{n-1\choose m}\frac{\frac
d{dx}S(n-1,m-1,x)}{S(n-1,m-1,x)^2}>0
\]
or 
\[
{S(n-1,m-1,x)^2}~>~
{n-1\choose m}\frac d{dx}S(n-1,m-1,x)~.
\]
To simplify this expression, which we want to show for
$n>m>0$, we equivalently show for $n>m\ge0$ that
\[
{S(n,m,x)^2}~>~
{n\choose m+1}\frac d{dx}S(n,m,x)
\]
which by (\ref{Sx}) says
\begin{equation}
\label{din}
\left(\sum_{i=0}^m{n-1-i\choose m-i}x\ttp{i} \right)^2
> 
{n\choose m+1}
\sum_{i=1}^mi{n-1-i\choose m-i}x\ttp{i-1} ~.
\end{equation}
In (\ref{din}), we have $x>0$ and (because $n-1\ge m$)
positive coefficients of $x\ttp m$ (and of higher powers
of~$x$) on the left hand side, whereas the highest power of
$x$ on the right hand side is $x\ttp{m-1}$.
Hence, it suffices to show that for $0\le i\le m-1$, each
coefficient of $x\ttp i$ on the left hand side in (\ref{din})
is at least as large as the coefficient of $x\ttp i$ on the
right hand side, that is, 
\begin{equation}
\label{coeff}
\sum_{k=0}^i{n-1-k\choose m-k}{n-1-i+k\choose m-i+k}
\ge
{n\choose m+1}(i+1){n-2-i\choose m-1-i}.
\end{equation}
There are $i+1$ summands on the left of (\ref{coeff}), so it
suffices to show that each of them, for $0\le k\le i$, fulfills
\begin{equation}
\label{sumd}
{n-1-k\choose m-k}{n-1-i+k\choose m-i+k}
\ge
{n\choose m+1}{n-2-i\choose m-1-i}.
\end{equation}
Because
\[
{n\choose m+1}
={n-1-k\choose m-k}\prod_{j=0}^k\frac{n-k+j}{m+1-k+j}
\]
and
\[
{n-1-i+k\choose m-i+k}
=
{n-2-i\choose m-1-i}\prod_{j=0}^k\frac{n-1-i+j}{m-i+j}~,
\]
(\ref{sumd}) is equivalent to 
\[
\prod_{j=0}^k\frac{n-1-i+j}{m-i+j}
~\ge~
\prod_{j=0}^k\frac{n-k+j}{m+1-k+j}
\]
which holds if for $0\le j\le k$ 
\[
\frac{n-1-i+j}{m-i+j}
~\ge~
\frac{n-k+j}{m+1-k+j}~,
\]
that is,
\begin{equation}
\label{1+}
1~+~\frac{n-m-1}{m-i+j}
~\ge~
1~+~\frac{n-m-1}{m+1-k+j}~.
\end{equation}
If $n=m+1$, then (\ref{1+}) holds as equality.
Otherwise, $n-m-1>0$, and (\ref{1+}) is equivalent to 
\[
m-i+j~\le~m+1-k+j
\]
or $k\le i+1$ which is true.
This proves the claim and thus (\ref{phat}), and the theorem.
\endproof

\section{Conclusions}
\label{s-conclude}
\def\pgame{\prec}

We have presented an inspection game that extends existing
models of $n$ time periods and $m$ inspections with a
general model of $k$ intended violations.
Each violation may have a different marginal reward to the
inspectee (which is therefore completely general) and a
proportional penalty when caught.
As shown in Theorem~\ref{t-gen}, this proportional penalty
is necessary and sufficient for applying the recursively
described game with full information to the game without
full information where the inspector in uninformed about the
inspectee's past actions in uninspected time periods, which
is a realistic condition.
We have studied three variants of this game:
A zero-sum game, a non-zero-sum game where both inspector
and inspectee get negative payoffs for a caught violation,
and a leadership game.
In the leadership game, the inspector commits to the same
mixed strategy but receives a higher payoff because the
inspectee acts legally as long as inspections remain (which
evidently requires commitment power of the inspector who
would otherwise not inspect in response).
We were able to give explicit solutions for these games by
proving nontrivial binomial identities and inequalities, as
in the proofs of Theorems \ref{t-main} and \ref{t-yeslegal}.

Further extensions of the model could involve inspections
with statistical errors of false alarms and non-detected
violations (as studied for a single intended violation 
in \cite{Ri1996}, and in a different model where time to
detection matters in \cite{AvCa2005}).
The fixed number $n$ of time periods could be replaced by a
random variable that counts ``suspicious events''.
Rather than a fixed number $m$ of inspections, an overall
frequency (with a public randomization device) could part of
the treaty for an inspection regime.
Also, there could be multiple inspectees, with different
targets for violations.

Such extensions pose without doubt new challenges for
analysis.
In practice, a new game of this sort is most likely not
solved explicitly, as in this paper, but with the help of
computer algorithms.
In fact, variations of the model investigated in 
\cite{vS1991} led the author to the study of algorithms for solving
extensive games, and the computationally efficient
``sequence form'' described in \cite{vS1996}.

A recursive game is a much more compact description than an
extensive game.
The inspection game $\Gamma(n,m,k)$ in (\ref{rec}) is
defined recursively in terms of the game values for simpler
games.
There are on the order of $n\cdot m\cdot k$ simpler games,
a polynomial number in $n,m,k$.
In contrast, the extensive game has an exponential number of
nodes, because every set of $m$ out of $n$ time periods that
the inspector inspects defines a different history of the
game.
However, from this history only the number of remaining
inspections matters, which is captured by the recursive
description.

Everett \cite{Ev1957} defined recursive games as stochastic games
where only absorbing states have nonzero payoffs.
The recursive games we consider could be put in this form by
awarding all payoffs (with the gains
$r_k,r_{k-1},\ldots$ to the inspectee for successful
violations) at the end of the game, which is possible
because the number $k$ of remaining violations is part of the
state description as $(n,m,k)$.
More importantly, in our games no state is ever revisited
during play because $n$ is decreased each time.
The game graph of state transitions is acyclic, but not a
tree.

Can such recursive games be equipped with nontrivial
information structures?
By Corollary~\ref{c-nonzinf}, the inspection game
$\hat\Gammaprime(n,m,k)$ without full information has the
same solution as the recursive game $\hat\Gamma(n,m,k)$ with
full information.
Here this is possible because the information that is
implicit in the recursive description does not matter,
which is due to the special payoff structure.
Otherwise the equilibrium strategy of the inspector depends
on~$k$.
In that case, can the equilibrium of the game with full
information be used to solve the game without full
information?
(This question was posed to the author by the late Michael
Maschler in 1991.)
An interesting area of future work could be models of
``small'' games like recursive games that allow for lack of
information, similar to information sets in extensive games,
and corresponding solution methods.  

%
%
%

\section*{Acknowledgments}
The author thanks Rudolf Avenhaus and Shmuel Zamir for their
interest in the unpublished precursor to this work in 
\cite{vS1991}, and the associate editor and a referee
for helpful comments, which led to Theorem~\ref{t-gen}.
Johannes H\"orner raised the question whether the
equilibrium of the non-zero-sum game is unique, which led to
Theorem~\ref{t-unique}.
Further thanks are due to participants at recent seminars
at LSE (London), Maastricht, Paris, Strathclyde (Glasgow),
and USC (Los Angeles).

\end{document}